


\magnification=1200

\vsize=7.5in
\hsize=5.6in
\tolerance 10000

\baselineskip 12pt plus 1pt minus 1pt
\pageno=0
\centerline{{\bf TIME TRAVEL?}\footnote{*}{This
work is supported in part by funds
provided by the National Science Foundation under grant \#PHY-88-04561 (SD)
and the U. S. Department of Energy (D.O.E.) under contract
\#DE-AC02-76ER03069 (RJ).}}
\vskip 24pt
\centerline{S.~Deser}
\vskip 12pt
\centerline{\it Department of Physics}
\centerline{\it Brandeis University}
\centerline{\it Waltham, Massachusetts\ \ 02254\ \  \ U.S.A.}
\vskip 12pt
\centerline{and}
\vskip 12pt
\centerline{R.~Jackiw}
\vskip 12pt
\centerline{\it Center for Theoretical Physics}
\centerline{\it Laboratory for Nuclear Science}
\centerline{\it and Department of Physics}
\centerline{\it Massachusetts Institute of Technology}
\centerline{\it Cambridge, Massachusetts\ \ 02139\ \ \ U.S.A.}
\vskip 1in
\centerline{\it In Memoriam}
\medskip
\centerline{\bf Gary Feinberg}
\vfill
\centerline{Extended version of talk presented at}
\centerline{`46 LNS 46}
\centerline{Cambridge, MA, May 1992}
\vfill
\centerline{ Typeset in $\TeX$ by Roger L. Gilson}
\vskip -12pt
\noindent BRX TH 334 \hfill June 1992

\noindent CTP\#2101
\eject
\baselineskip 24pt plus 2pt minus 2pt
To travel into the past, to observe it, perhaps to influence it and correct
mistakes of one's youth, has been an abiding fantasy of mankind
for as long as we have
been aware of a past.  Here are described some recent scientific
investigations on this topic.

Before the twentieth century, time travel was discussed only in works of
fiction;
among innumerable instances, best known are surely the
novels of Mark Twain and H. G. Wells.  The latter marks a transition:  Wells,
a graduate of London's Imperial College of Science and Technology, couched
his novel {\it The Time Machine\/} in scientific language, giving us an early
work of science fiction.

After 1900, special relativity made scientific discussion of time
machines possible.  The question may be posed in this
way: it makes perfectly good sense to speak about
 travel that returns to the same
point in {\it space\/}.  But Einstein and Minkowski
tell us that space and time are
equivalent, so after a  journey can we return
to our starting position in {\it time\/}?
Here the answer is well-known: the space-time of
special relativity is flat and rigid; while time travel into the past
is indeed possible, it requires faster-than-light velocities.  This is seen as
follows.  The Lorentz transformation law of space $(\Delta x)$ and time
$(\Delta t)$ intervals in some initial (inertial) reference frame
into corresponding quantities (designated by an
overbar) in another frame, moving relative to the original one with frame
velocity $v$ along say the $x$-axis, is given by
$$\eqalignno{\overline{\Delta t}
&= {1\over \sqrt{1 - v^2/c^2}} \left( \Delta t -
{v\over c^2}\Delta x\right) &(\hbox{1a}) \cr
\overline{\Delta x} &= {1\over \sqrt{1 - v^2/ c^2}} \left( \Delta x - v
\Delta t\right)\ \ . &(\hbox{1b}) \cr}$$
Since the
velocity $u$ of a moving object in the original frame is $\Delta x/\Delta t$,
we also have
$$\overline{\Delta t} = {1\over \sqrt{ 1 - v^2/ c^2}} \Delta t \left( 1 -
{uv\over c^2}\right)\ \ . \eqno(2)$$
For  $\overline{\Delta t}$ to remain real, the frame velocity $v$ must not
exceed the velocity of light $c$, but $\overline{\Delta t}$ can have a sign
opposite to  $\Delta t$ if
$${u/ c} > {c/ v} > 1 \ \ .\eqno(3)$$
One can think of this result as a continuation of the familiar time-dilation
story: the faster one travels starting from rest, the slower flows time; time
stands still for travel at the velocity of light and runs backward once $c$
 is exceeded.

A time machine is now constructed in the following manner.
When a faster-than-light
object --- called a {\it tachyon\/} --- is emitted with $u>c$
by a source, moving backwards with speed
$v<c$ and satisfying (3), the tachyon will arrive at its goal a time
$\left|\overline{\Delta t}\right|$
before it was sent out; then if returned by
a similarly backward-moving source, it will arrive at its point of origin at
$2\left|\overline{\Delta t}\right|$ before emission.
However, there are no known tachyons, and
 time machines cannot be constructed with the
physics and engineering possibilities provided by special relativity.
Nevertheless, it should be stressed that there is no {\it logical\/}
prohibition
against exceeding light velocity.  Indeed a hypothetical world in which
tachyons exist is physically consistent.  Although
tachyons have been looked for
experimentally none have ever been found; see Ref.~[1] for a nice review of
the entire subject.

But then we come to Einstein's general relativity, within which space-time
geometry is
no longer fixed; indeed it can take all kinds of unexpected configurations,
depending on the matter content of the universe.  In
particular there can be geometries containing paths along which one can
 travel into the past with
velocity {\it less\/}
than that of light --- such paths are called {\it closed time-like
curves\/}.  The first solution to Einstein's theory with closed time-like
curves
was obtained in 1949
by G\"odel and it permits construction of time machines.$^2$

G\"odel solved the Einstein gravity equations of general relativity
$$G_{\mu\nu} + {\Lambda\over 2} g_{\mu\nu} =
{8\pi G\over c^4} T_{\mu\nu} \ \ .\eqno(4)$$
The left side is geometrical; it contains
the Einstein tensor, $G_{\mu\nu}$, formed from
the metric tensor $g_{\mu\nu}$ in which is encoded
the geometry of space-time, together with a
cosmological term.  The right side describes matter, $T_{\mu\nu}$ being
its local energy-momentum distribution; matter determines geometry through
Eq.~(4). (Newton's $G$ is the appropriate dimensional
proportionality constant.)

G\"odel took $T_{\mu\nu}$ to be a space-time constant, not
vanishing only in its time-time component
(energy density),
$$T_{00} = {c^4\over 8\pi G} \Lambda >0\ \ .\eqno(5)$$
The metric tensor that then solves Einstein's equations leads to the
space-time interval
$$\eqalign{ds^2 &= g_{\mu\nu}\, dx^\mu\, dx^\nu \cr
&= \left(c\, dt - \sqrt{{2\over\Lambda}}\left(\cosh\sqrt{\Lambda}\,
r - 1 \right)d\theta\right)^2 -
dr^2 -{1\over\Lambda}\ \sinh^2 \sqrt{\Lambda}\,r
\ d\theta^2 - dz^2
\ \ ,\cr}\eqno(6)$$
where $r$, $\theta$ are planar circular coordinates, with $\theta=0$ and
$2\pi$ identified, and there is no
interesting structure in the $z$-direction.  A curve $x^\mu(\tau)$
is closed and
time-like if both $x^\mu(0) = x^\mu(1)$
(closed) and $\left( ds/d\tau\right)^2 = g_{\mu\nu} \left( dx^\mu/d\tau\right)
\left( dx^\nu/d\tau\right) >0$ (time-like).
It is therefore clear that a circular path in
the G\"odel universe for which $t$, $r$ and $z$ remain constant, while $\theta$
varies from $0$ to $2\pi$, is closed and time-like provided
$\cosh\sqrt{\Lambda}\, r>3$, {\it i.e.\/}, $r>{2\over \sqrt{\Lambda}}\ln
(1+\sqrt{2}\,)$.

This result caused great puzzlement, because first of all
there is no evidence for time travel ---
we know of no visitors from the future, and second our classical notions of
causality (which to be sure are already challenged by quantum mechanics)
prejudice us against considering geometries with closed time-like curves,
where effects precede causes.  Here is Einstein's reaction$^3$
\smallskip
{\noindent\narrower ``\sl Kurt G\"odel's [time machine solution raises] the
problem [that] disturbed me already at the time of the building up of the
general theory of relativity, without my having succeeded in clarifying it.
$\cdots$ It will be interesting to weigh whether these [solutions] are not to
be excluded on physical grounds.''\smallskip}
\smallskip
\noindent A more recent comment is by Hawking.$^4$
\smallskip
{\noindent\narrower\sl ``G\"odel presented a $\cdots$ solution [which] was the
first to be discovered that had the curious property that in it it was
possible to travel into the past.  This leads to paradoxes such as `What
happens if you go back and kill your father when he was a baby?'  It is
generally agreed that this cannot happen in a solution that represents our
universe, but G\"odel was the first to show that it was not forbidden by the
Einstein equations.  His solution generated a lot of discussion of the
relation between general relativity and the concept of causality.''\smallskip}
\smallskip
Upon further reflection it comes as no surprise that Einstein's general
relativity allows closed time-like curves: in Einstein's theory geometry is
determined from matter by Eq.~(4) whose schematic form is
$$\left.\matrix{\hbox{space-time}\cr\hbox{geometry}\cr}\right\}
\Longleftarrow \left\{
\matrix{\hbox{distribution}\cr\hbox{of matter}\cr}
 \right.\ \ .$$
But Einstein's equations may be read in the opposite direction:
pick an interesting geometry --- no matter how strange --- and in particular
one containing closed time-like curves, then determine the (unphysical?)
 matter
distribution that engenders it, thereby ``finding'' a time machine solution,
{\it i.e.\/}, redirecting the arrow above:
$$\left. \matrix{\hbox{time machine}\cr\hbox{geometry}\cr}\right\}
\Longrightarrow \left\{\matrix{\hbox{peculiar matter}\cr\hbox{``unphysical?''}
\cr}\right.\ \ .$$

Our ``defense'' against these solutions and against the paradoxes
they entail is to assert that the exotic matter distributions supporting time
machines are unphysical.  Thus G\"odel's universe requires a constant and
uniform energy density, which clearly is unphysical.  The time machines studied
by Thorne and his colleagues$^5$ at CalTech
make use of wormholes, another exotic and presumably
unphysical form of matter in which a narrow channel --- the wormhole ---
connects distant regions of space-time and is threaded by a closed time-like
curve.$^6$

The reason for current interest in time travel ideas
derives from the recent realization that infinitely long and arbitrarily thin
cosmic strings can support closed time-like curves. Cosmic strings
are hypothetical, but
entirely physical structures that may have survived from a cosmic phase
transition, and that may
even be responsible for the present-day large scale structure in
the universe.$^8$
Only completely conventional physical ideas are relied upon in
cosmic string speculations;  cosmologists make use of cosmic strings in model
building, and astrophysicists occasionally report sightings, although thus
far evidence is inconclusive.  Nevertheless, although infinite length and
arbitrary thinness are idealizations, one would not view such cosmic strings
as unphysical.  Therefore, if they indeed give rise to closed time-like
curves, there is something new that physicists must confront.

To recognize string-generated closed time-like curves, we need to record the
space-time that is produced by a cosmic string. We use coordinates in which an
infinitely long and thin cosmic string lies
 along the $z$-axis through the origin.  Its mass per
unit length is $m$ and we also endow the string with intrinsic spin $J$ per
unit length.  The cross-sectional area
is assumed to be vanishingly small,
so the mass density is proportional to a two-dimensional spatial $\delta$-
function, while the spin density is more singular, being produced by a
momentum density proportional to derivatives of $\delta$.

Solving Einstein's equation (4) (with vanishing cosmological constant
$\Lambda$) for a stationary string,
carrying the above mass and spin distributions,
produces a metric described by the
line element
$$\eqalign{ds^2 &= g_{\mu\nu} dx^\mu\, dx^\nu \cr
&= \left(c\, dt +{4\over c^3} GJd\theta\right)^2 - dr^2 - \left( 1 - {4\over
c^2} Gm\right)^2 r^2\, d\theta^2 - dz^2 \cr}\ \ .\eqno(7)$$
Indeed this two-parameter ($m,J$) metric tensor is the general time-independent
solution to (4) outside any matter distribution lying in a bounded region on
the plane and having cylindrical symmetry.
As in the G\"odel line element (6), there is no structure in the $z$-direction,
while in the perpendicular ($r,\theta$) plane, with $\theta=0$ and
$2\pi$ identified, the non-trivial geometry
supports closed time-like curves when $J$ is non-vanishing: take constant $t$
and $r$ and describe a circle  ($\theta$ ranging from 0 to $2\pi$)
with sufficiently small radius
$$r< {4 GJ\over c^3-4Gmc}\ \ .\eqno(8)$$

By changing coordinates one can hide the presence of the string and make the
line element appear locally Minkowskian: with the definitions
$$\tau= t +{4 \over c^4} GJ
\theta\ \ ,\qquad \varphi = \left(1-{4\over c^2} Gm\right)
\theta\ \ ,\eqno(9)$$
(7) becomes
$$ds^2 = (c\,d\tau)^2 - dr^2 - r^2\, d\varphi^2 - dz^2\ \  .\eqno(10)$$
Now, however, the ranges of these
flat-looking coordinates are unconventional:   that of
$\varphi$ is diminished
to $\left(1-4Gm/c^2\right)2\pi$, while
 the new time variable $\tau$, rather than
flowing in a smooth and linear fashion, jumps by $8\pi GJ/c^4$ whenever the
string is circumnavigated, owing to the identification of $\theta=0$ and
$2\pi$.  The defect in the angular range turns the spatial plane into the
surface of a cone with deficit angle determined by $m$, while the
time-helical structure, where pitch is proportional to intrinsic spin $J$,
is responsible for the closed time-like curves~(8).

There is another entirely equivalent framework for understanding the
geometry produced by our cosmic string:
In the idealization of infinitely long and thin strings, there is no structure
along the $z$-axis, hence we can suppress that direction altogether.  Then
the theory becomes ($2+1$)-dimensional ``planar'' gravity, governed by the
same Einstein
equations (4) but in this reduced space. Cosmic string sources now become
``point-particles'' at the locations where the strings pierce the $z=0$ plane.

Einstein's gravity theory in $(2+1)$ dimensions is a much-studied model,$^9$
both
for pedagogical reasons --- one hopes that the dimensional reduction effects
sufficient simplification to permit thorough analysis, while still retaining
useful content to inform the physical $(3+1)$-dimensional problem --- and also
for practical calculation --- as explained above, idealized cosmic
strings are effectively described by $(2+1)$-dimensional gravity. [Analogously,
motion of charged particles
in magnetic fields that are constant along the $z$-axis is effectively
governed by planar dynamics, as in Landau theory, quantum Hall effect and
perhaps high-$T_c$ superconductivity.]
We shall use
the name {\it cosmon\/} to refer uniformly to infinite cosmic strings in space
and to particles on the plane.

A simplifying feature of the lower-dimensional interpretation (always
without cosmological constant, although it can be included) is
that space-time is locally flat whenever sources ($T_{\mu\nu}$) are absent.
The reason for this is that the Riemann  curvature
tensor, which completely determines geometry, is linearly related to the
Einstein tensor
$G_{\mu\nu}$, in (and only in) three-dimensions (they are each other's
double duals).
Thus at any point where there are no sources, $G_{\mu\nu}$ vanishes by (4),
and consequently the three-dimensional
Riemann tensor also vanishes: space-time is Minkowskian.  Moreover, when
sources are point-cosmons, the space-time is flat everywhere, except at these
points, and this is why it is possible to transform the line element
(7) to the flat one (10) almost everywhere; only the
unconventional range and jump
properties of the coordinates remind us that there are sources somewhere.
Indeed these coordinate defects contain all the physical information about
sources that is accessible in regions outside them.$^{10}$

Further examination of the geometry generated by cosmons
shows that the mass of any
individual one ($=$ mass per unit length of a cosmic string) must
not exceed $c^2/4G$, so that the deficit angle not exceed $2\pi$ ($m=c^2/4G$
corresponds to the cone closing into a cylinder).
  The {\it total\/} mass of an {\it assembly\/} of static cosmons,
each with acceptable mass, may exceed $c^2/4G$, but then it must precisely
equal
$c^2/2G$ for space to be non-singular, and space is necessarily closed.$^{10}$
[This is a consequence of the fact that
the total mass of static sources
in ($2+1$)-dimensional gravity is proportional to the Euler
number of the 2-space.]  Note in particular: sources, which individually, {\it
i.e.\/} {\it locally\/}, are acceptable ($m<c^2/4G$) can give rise to
configurations that are {\it globally\/} unacceptable
($c^2/4G<\sum_mm\not=c^2/2G$); this lesson will be essential in what follows.
\pageinsert
\vskip 3.25in
\centerline{\it Fig.~1:\quad Space of a spinless cosmon.}
\vfill
\centerline{\it Fig.~2:\quad Space-time of a spinless cosmon.}\endinsert

In Figs.~1--3 we depict the space and the space-time of a single cosmon using
flat coordinates with unconventional ranges.  In Fig.~1 the cosmon is
spinless; the three-dimensional space is structureless along the $z$-direction
(direction of the string), in the perpendicular plane an angular region,
of magnitude to $8\pi Gm/c^2$, is excised and the edges identified ---
the space is a cone.  In Fig.~2, the $z$-axis is suppressed, the cosmon is a
point-particle and the time axis as well as the world line of the stationary
cosmon are indicated; since there is no spin, time flows smoothly from
$-\infty$ to $\infty$.  Figure 3 depicts the previous situation, but now the
cosmon also
carries spin and time acquires a helical structure, as explained above.
 For more cosmons, the space is an assembly of cones, joined smoothly,
as is possible provided each of their deficit angles and the sum is
suitably bounded by the limits given above.

\midinsert
\vskip 3.25in
\centerline{\it Fig.~3:\quad Space-time of a spinning cosmon.}\endinsert

Another useful way of describing the space-time created by a cosmon is to
notice that after the cosmon is circumnavigated, the flat space-time
description
requires an additional rotation in space and a jump in time, which effect the
coordinate identifications that encode the presence of a massive, spinning
cosmon.  These identifications may be represented in terms of
a Poincar\'e transformation (spatial rotation or Lorentz
boost and space-time translation) on the three-dimensional space-time (since
the invariance group of flat space-time is the Poincar\'e group),
$$\overline{x}^\mu = M^\mu{}_\nu x^\nu + a^\mu\ \ .\eqno(11)$$
Here $\overline{x}$ and $x$ are the identified three-vectors,
$M^\mu{}_\nu$
is a Lorentz transformation, specialized for the single static source
to a spatial rotation through the
angle $8\pi Gm/c^2$, and $a^\mu$ is the jump in coordinates, specialized to a
jump of $8\pi GJ/c^4$ in the time component. A
  general cosmon configuration, not necessarily a single cosmon at
rest but also a moving one or indeed an assembly of cosmons, is also
described by such a Poincar\'e identification, with $x$
and $\overline{x}$ referring to points exterior to the assembly.  For example,
 a spinless cosmon
moving with velocity ${\bf v}$ through the point ${\bf x_0}$ leads to the
identification
$$\overline{x} = B^{-1}_{\bf v} M_m B_{\bf v} (x-x_0) + x_0 \ \ ,\eqno(12)$$
where $B_{\bf v}$ is the Lorentz boost that brings the cosmon to rest
and $M_m$ is the mass-determined rotation.
Equation~(12) is merely the statement that the coordinate $B_{\bf v} (x-x_0)$
is identified as in (11), with $a$ vanishing when there is no spin.

As already remarked, owing to the time-helical structure that is required by
the Einstein equation,
a spinning ($J\not=0$) cosmon gives rise to closed time-like curves.
Now {\it intrinsic\/} spin attached to a cosmic string may still be deemed
unphysical (like any classical spin involving derivatives of
$\delta$-functions),
so we need not worry about closed time-like curves supported by locally
spinning cosmons.  However, one may ask whether {\it two\/} or more spinless
cosmons, moving relative to each other and thus also carrying {\it orbital\/}
angular momentum, can still
support closed time-like curves.  This questions was answered affirmatively by
Gott at Princeton University.$^{11}$  He found that two spinless cosmons,
each moving faster than a critical but subluminal ($v<c$) velocity,
do indeed support closed time-like curves.

In fact the question of possible closed time-like curves in a many-spinless
cosmon universe had been posed even earlier and answered negatively
in the original investigation that first exhibited closed time-like curves in
the presence of spinning cosmons.$^{10}$  Briefly, the reasoning there
 relied on
the expectation that the perfectly good Cauchy development of two freely
moving cosmons from $t=-\infty$ should not be spontaneously destroyed at some
finite time, when they approach each other, and then be reinstated when they
diverge as $t\to+\infty$.  Resolution of the apparent
contradiction between this expectation and Gott's result
is the most recent development in the time machine story.
\midinsert
\vskip 3.25in
\centerline{\it Fig.~4:\quad Two-cosmon time machine.}\endinsert

The disposition of cosmons that support closed time-like curves is portrayed
in Fig.~4.  Two spinless cosmons, each with equal (for simplicity) masses $m$
are located at the center of the
figure, and the excised wedges point up and down.  Consider now the light path
from $A$ to $B$.   It is credible that a path
surrounding the cosmons and taking advantage of the ``missing'' space can be
shorter than the direct, straight line $AB$.  Then cosmons moving in
opposite directions with speed $v$, in a manner already described above
in connection with the tachyonic time machine, create closed time-like curves,
provided $v$ exceeds a critical value $v_{\rm critical}$, which
nevertheless is less than $c$, so that neither cosmon is tachyonic.  Gott
shows that
$${v\over c}>{v_{\rm critical} \over c} = \cos\left({4\pi Gm\over c^2}\right)
< 1\ \ .\eqno(13)$$

This result, in what purports to be a physically acceptable situation,
produced intense interest (not only in the physics community, as is seen from
the appended list of semi-popular accounts$^{12}$)
since for the first time time-travel appeared to be consistent with
unexceptional physical
principles, awaiting only development of engineering possibilities.  Fantastic
applications obviously come to mind and were discussed, while at the same time
some pitfalls inherent in commercial development were noted in a nationally
syndicated newspaper feature, published the same day as the analysis in
Ref.~[13] and reproduced in Fig.~5.
\midinsert
\vskip 3.2in
\centerline{\it Fig.~5:\quad Pitfall in commercial development of time
machines.}
\endinsert
But there is a catch: since one needs a {\it pair\/}
of moving strings, one may ask
what is their combined energy and momentum.  Now energy and momentum are
conserved quantities that arise from invariance under time- and
space-translation.  However, when space-time is as globally complicated ---
``sewn-together'' cones --- as it becomes
in the presence of strings, the addition rules for combining energy and
momentum become non-trivial and non-linear.  In particular, even though each
of the two cosmons is moving sufficiently slowly to be non-tachyonic,
as soon as the velocity of each exceeds the critical value (13)
needed to support a time machine, their center-of-mass becomes tachyonic.

This is seen as follows.  We have already remarked that a general cosmon
distribution is characterized by an element of the Poincar\'e group, as in
(11).  Concentrating now on the ($2+1$)-dimensional Lorentz transformation
matrix $M$, we recall that it may be presented as the exponential of three
generator matrices $\Sigma_\mu$,
$$M = \exp\left\{ {8\pi\over c^3} GP^\mu \Sigma_\mu\right\}\ \ .\eqno(14)$$
[This is analogous to presenting a three-dimensional spatial
rotation matrix as the
exponential of angular momentum matrices.]  The three-vector $P^\mu$  defines
the energy-momentum of a cosmon assembly.  For
example, when a single cosmon is at rest, $M$ is a pure rotation, which
corresponds to
$${\bf P} = 0\ \ ,\qquad P^0 = mc\ \ ,\qquad \Sigma_0 \equiv \left(
\matrix{ 0 & \phantom{-}0 & 0 \cr0 & \phantom{-}0 & 1 \cr 0 & -1 & 0 \cr}
\right)\ \ ,\eqno(15)$$
where the rows and columns are labelled by $(t,x,y)$.
For the cosmon assembly to be non-tachyonic, {\it i.e.\/} physically
constructible, we must require
$$P^\mu P_\mu \equiv m^2c^4>0\ \ .\eqno(16)$$

Now Gott's time machine makes use of two equal-mass ($m$) cosmons, moving in
opposite directions with speed $v$.  Suitably extending (12), we see that the
resulting Lorentz identification that describes it is
$$M_{\rm total} = B^{-1}_{\bf v} M_m B_{\bf v}B^{-1}_{\bf -v}
M_m B_{-{\bf
v}} = B^{-1}_{\bf v} M_m B^2_{\bf v} M_m B^{-1}_{\bf v}
\eqno(17\hbox{a})$$
with total energy and momentum defined by
$$M_{\rm total} = \exp\left\{{8\pi G\over c^3}
P^\mu_{\rm total}\Sigma_\mu\right\}\ \ .\eqno(17\hbox{b})$$
A straightforward calculation yields the formula for the total mass, $m_{\rm
total}$,
$$\sin\left({2\pi Gm_{\rm total}\over c^2}\right)
= {1\over \sqrt{1-v^2/c^2}} \sin\left({4\pi Gm\over c^2}\right)\ \ ,\eqno(18)$$
which for small $Gm/c^2$ recovers the special relativistic result
$$m_{\rm total} = {2m\over \sqrt{1-v^2/c^2}}\ \ .$$
For real $m_{\rm total}$, the left side of (18) is less than unity, so
we must have
$${v\over c} < \cos\left( {4\pi Gm\over c^2}\right)\ \ .\eqno(19)$$
This is precisely {\it opposite\/} to Gott's criterion (13) for the presence of
closed time-like curves!
Thus we learn that when
(13) is satisfied, {\it each\/} cosmon travels with a velocity
below that of light, but the {\it total\/} system  possesses a tachyonic
energy.
It follows that only tachyons can give rise to Gott's
sufficiently rapidly moving pair
of cosmons.$^{13}$
In the absence of such tachyonic sources, energy is unavailable
to produce the rapidly moving cosmons.
Another perspective is given by formula (18).  We see that when $v=0$,
$m_{\rm total} = 2m$.  As $v$ increases from zero, $4Gm_{\rm total}/c^2$
increases and approaches unity when $v$ reaches $v_{\rm critical}$,
corresponding to a deficit angle $2\pi$ in the
surrounding space.
As the critical velocity is exceeded, the deficit
angle surpasses $2\pi$, and the system becomes unphysical.

Thus the two opposite claims$^{10,\,11}$ about time machines are
reconciled:$^{13}$  just like
the special relativistic time machines, which could only be constructed if
tachyons exist --- but they do not --- so also the cosmic string time machines
require tachyonic center-of-mass velocities, and cannot be produced in the
absence of tachyons, that is in our world.

Even without mentioning the tachyonic center-of-mass, we can give an
equivalent geometrical description of the unphysical nature of Gott's system:
Its exterior geometry necessarily contains closed time-like curves at
spatial infinity.  Existence of a time machine during
that portion of the system's evolution when the cosmons are sufficiently
close requires that closed time-like curves be also present at the boundary of
space for {\it all\/} time.  Space-times
with this property have long been known in
classical gravity, where they are produced by unusual identifications of space
and time.  An example is Misner space,$^{14}$ and one can show that the
coordinate identifications by means of the Lorentz transformation (17a)
when $v>v_{\rm critical}$
give  rise to such a space.  Moreover, the {\it effective\/}
energy-momentum tensor responsible for the exterior Gott metric is
tachyonic, with energy density replaced by spatial stress.

The above analysis was confirmed in further investigations on  the
obstacles to constructing Gott's time machine.  Carroll,
Farhi and Guth$^{15}$ showed that it cannot be manufactured
as a subsystem of a normal time-like matter assembly in an {\it open\/}
universe.  In a universe that is spatially {\it closed\/},
owing to a mass distribution
with critical magnitude for closing space ($\sum_mm=c^2/2G$), a pair of
cosmons traveling with velocity exceeding $v_{\rm critical}$
{\it can\/} be produced;$^{15}$ nevertheless
a time machine cannot be built here either because the universe dramatically
self-destructs just before the cosmons get close enough to engender a closed
time-like curve --- there is not enough time to travel into the past.$^{16}$

While we are reassured,
at least as concerns this most
recent string-inspired attempt, that causality will not be violated by time
travel, further questions remain
and motivate further study.  One wonders whether
one can {\it a priori\/} and once-and-for-all characterize those mass
distributions that {\it mathematically\/} support time travel and therefore
should perhaps be deemed unphysical.  Indeed, Hawking$^7$ has encapsulated
these ideas in his ``Chronology Protection Conjecture'' that ``the laws of
physics do not allow the appearance of closed time-like curves.''
But establishing this conjecture requires
knowing why the laws of physics prohibit certain forms of matter.  That this
question is difficult to answer is well-illustrated by the cosmon time machine:
each individual cosmon is a perfectly acceptable subluminal and physical form
of matter --- but fitting exactly two rapidly moving
cosmons into a non-singular universe is
impossible.  Evidently global, rather than local considerations come into the
definition of ``physical.''

However, it should be remarked that as paradoxical as time-travel may appear
(``killing one's infant father,'' {\it etc.\/}) we can be assured that
mathematical analysis, correctly carried out, will provide a
consistent physical picture --- which of course need not
correspond to reality if conditions for time-travel cannot be physically
met.  This had
been already established for tachyonic effects within special relativity.$^1$
Recent investigations of physical processes in the presence of closed
time-like curves, like those generated by spinning strings$^{17}$ or by other
mechanisms$^{18}$ confirm that unexpected but not self-contradictory
results are found, {\it e.g.\/}, perturbative violation of unitarity in
interacting quantum field theories.

We study paradoxical aspects of physics --- like time machines
--- to illuminate our understanding of
fundamental issues.  For example, the
very concept of time remains mysterious in physical theory:
What gives time its observed
direction?  Indeed what gives it a definition, in the first place, since
general relativity requires space-time to be determined by
matter?  It should be apparent from the foregoing that here there is still much
to learn.
\vfill
\eject
\centerline{\bf REFERENCES }
\medskip
\item{1.}G. Feinberg, {\it Scientific American\/} {\bf 222}(2), 68 (1970).
\medskip
\item{2.}K. G\"odel, {\it Rev. Mod. Phys.\/} {\bf 21}, 447 (1949).
\medskip
\item{3.}A. Einstein, in {\it Albert Einstein: Philosopher--Scientist\/}, P.
Schilpp, ed. (Tudor, New York, 1957).
\medskip
\item{4.}S. Hawking, in {\it K. G\"odel, Collected Works\/}, Vol.~II, S.
Feferman, ed. (Oxford University Press, New York, 1990).
\medskip
\item{5.}M. Morris and K. Thorne, {\it Am. J. Phys.\/} {\bf 56}, 395
(1988).
\medskip
\item{6.}Specific, technical objections can also be raised to
wormhole-supported
time machines.  The narrowness of the channel would prevent anything of
significant size from squeezing through its opening.  Moreover, Hawking$^7$
has argued that quantum fluctuations around a wormhole generate infinite
energy and stress, thereby closing it, but the subject is still being
investigated.
\medskip
\item{7.}S. Hawking, in {\it The Sixth Marcel Grossmann Meeting on General
Relativity\/}, H. Sato, ed. (World Scientific, Singapore, 1992).
\medskip
\item{8.}For a review, see A.~Vilenkin, {\it Physics Reports\/} {\bf 121}, 263
(1985).
\medskip
\item{9.}For reviews, see R. Jackiw, in {\it Relativity and Gravitation:
Classical and Quantum\/}, J. D'Olivio, E. Nahmad-Achar, M. Rosenbaum, M. Ryan,
L. Urrutia and F. Zertuche, eds. (World Scientific, Singapore, 1991), and
in {\it The Sixth Marcel Grossmann Meeting on General
Relativity\/}, H. Sato, ed. (World Scientific, Singapore, 1992).
\medskip
\item{10.}S. Deser, R. Jackiw and G. 't~Hooft, {\it Ann. Phys.\/} (NY) {\bf
152}, 220 (1984).
\medskip
\item{11.}J. Gott, {\it Phys. Rev. Lett.\/} {\bf 66}, 1126 (1991).
\medskip
\item{12.}{\it Time\/}, 13 May 1991; {\it Science News\/}, 28 March 1992;
{\it New Scientist\/}, 28 March 1992; {\it Discover\/}, April 1992;
{\it Science\/}, 10 April 1992.  Only the {\it New Scientist\/} and {\it
Science\/} articles convey accurate information.
\medskip
\item{13.}S. Deser, R. Jackiw and G. 't~Hooft, {\it Phys. Rev. Lett.\/} {\bf
68}, 267 (1992), and the last paper in Ref.~[9].
\medskip
\item{14.}C. W. Misner,
in {\it Relativity Theory and Astrophysics I: Relativity and
Cosmology\/}, J. Ehlers, ed. (American Mathematical Society, Providence,
1967).
\medskip
\item{15.}S. Carroll, E. Farhi and A.
Guth, {\it Phys. Rev. Lett.\/} {\bf 68}, 263, (E) 3368
(1992); see also D. Kabat, MIT
preprint CTP\#2034 (November 1991).
The claim in the first of these papers that
time machines can be constructed in closed universes was
shown to be false in Ref.~[16].
\medskip
\item{16.}G. 't~Hooft, {\it Class. Quant. Grav.\/} {\bf 9}, 1335 (1992).
\medskip
\item{17.}P. Gerbert and R. Jackiw, {\it Comm. Math. Phys.\/} {\bf 124}, 495
(1988).
\medskip
\item{18.}J. Friedman, M. Morris, I. Novikov, F. Echeverria, G. Klinkhammer,
K. Thorne and U. Yurtsever, {\it Phys. Rev. D\/} {\bf 42}, 1915 (1990); D.
Deutsch, {\it Phys. Rev. D\/} {\bf 44}, 3197 (1991); J. Friedman, N.
Papastamatiou and J. Simon,  University of Wisconsin--Milwaukee preprint
WISC-MIL-91-TH-17 (1991); J. Hartle, UC Santa Barbara preprint, UCSBTH-92-04
(1992);
D.~Boulware, University of Washington preprint UW/PT-92-04 (1992).
\par
\vfill
\end